\documentclass[twocolumn,portuges,british,showpacs,preprintnumbers]{revtex4}
\usepackage{ae,aecompl}
\usepackage[T1]{fontenc}
\usepackage[latin9]{inputenc}
\usepackage{units}
\usepackage{amsmath}
\usepackage{amssymb}
\usepackage{graphicx}

\makeatletter
\@ifundefined{textcolor}{}
{%
 \definecolor{BLACK}{gray}{0}
 \definecolor{WHITE}{gray}{1}
 \definecolor{RED}{rgb}{1,0,0}
 \definecolor{GREEN}{rgb}{0,1,0}
 \definecolor{BLUE}{rgb}{0,0,1}
 \definecolor{CYAN}{cmyk}{1,0,0,0}
 \definecolor{MAGENTA}{cmyk}{0,1,0,0}
 \definecolor{YELLOW}{cmyk}{0,0,1,0}
}

\usepackage{dcolumn}

\usepackage{bm}

\makeatother

\usepackage{babel}

\makeatother

\usepackage{babel}

\usepackage{hyperref}

\makeatother

\usepackage{babel}
\begin{document}

\title{Generation of entangled light via dynamical Casimir effect}

\selectlanguage{portuges}%

\author{A. Guerreiro}

\email{ariel@fc.up.pt}

\selectlanguage{portuges}%

\affiliation{INESC TEC (formerly INESC Porto), Rua do Campo Alegre, 687, Porto,
Portugal}

\affiliation{Departamento de F\'{\i}sica e Astronomia, Faculdade Ci\^{e}ncias Universidade
do Porto, 687 4169-007, Porto, Portugal}

\author{P. Alcino}

\affiliation{INESC TEC (formerly INESC Porto), Rua do Campo Alegre, 687, Porto,
Portugal}

\affiliation{Departamento de F\'{\i}sica e Astronomia, Faculdade Ci\^{e}ncias Universidade
do Porto, 687 4169-007, Porto, Portugal}

\selectlanguage{british}%

\date{\today}
\begin{abstract}
This paper addresses the excitation of vacuum fluctuations of the
electromagnetic field through periodic modulations of a refractive
index and the possibility of using entanglement as a distinctive marker
of the quantum nature of the phenomenon. It introduces a lossy environment
and analyses its implications on the possibility of generating such
an effect and measuring entanglement, concluding that it is not entirely
destructive when the produced particles share the same environment.

\end{abstract}
\maketitle
\textbf{\emph{Introduction.}} The dynamical Casimir effect (DCE) consists
basically in applying time varying boundary conditions or refractive
index modulations to the electromagnetic vacuum in order to excite
real particles out of the zero\textendash{}point fluctuations (ZPF)
\cite{10.1063/1.1665432,PhysRevA.53.2664,PhysRevA.49.433,1742-6596-21-1-025}.
Even though this is probably the most direct method to probe and enhance
ZPF, detecting it in optical cavities is not easy, albeit there being
some promising attempts in that direction \cite{1742-6596-161-1-012027,PhysRevLett.105.203901}
and observations in analogue experiments \cite{Wilson2011}. The obstacles
have mainly been attributed to the difficulty in producing intense
and fast enough optical perturbations \cite{1402-4896-82-3-038105,1742-6596-161-1-012027}
and to technical aspects of the experiment (cavity losses, detector
sensitivity, etc.) capable of measuring the scarce radiation produced
\cite{PhysRevA.58.4147,1402-4896-82-3-038105}. Finite temperature
has two added consequences: it provides thermal photons which seed
the DCE, thus masking the ZPF, and produces decoherence and losses
via the thermalisation of the optical cavity by the environment. In
fact, previous works have shown that a seed thermal state greatly
improves the photon production \cite{PhysRevLett.84.1882}. However,
as it will be demonstrated here, this amplification of thermal radiation
is also predicted with a classical model, rendering the DCE at finite
temperatures, in practice, a classical effect.

The key aspect of zero temperature DCE is that, unlike thermal fluctuations,
ZPF are coherent and the parametric amplification produced by a time
varying optical cavity preserves this coherence. For example, in an
alternative DCE scheme, which considers a medium with a time varying
refractive index in the absence of any spatial boundary, known as
time refraction (TR), the parametric amplification of ZPF generates
pairs of counter propagating photons which are maximally entangled
and exit the medium separately \cite{PhysRevA.83.052302}. For DCE
in linear cavities the counter propagating photons reflect into each
other, rendering them indistinguishable and thus impeding the use
of entanglement as a marker of quantum character. However, this difficulty
can be circumvented using different geometries.

Previous work in the problem considered the enhancement of the output
radiation due to thermal seeding \cite{PhysRevLett.84.1882} and the
impact of losses \cite{PhysRevA.58.4147,PhysRevA.80.023814,1742-6596-161-1-012027}
(phase damping, dissipation, Markovian baths, etc.) in linear optical
cavities. At finite temperature ($\mathcal{T}>0$) the losses in the
cavity do not force the cavity to the vacuum state but to the thermal
state of the cavity which is in equilibrium with the environment.
This provides constant thermal reseeding of the cavity (not just at
the initial moment) masking even further any mark of ZPF. 

\begin{figure}
\begin{centering}
\includegraphics[scale=0.2]{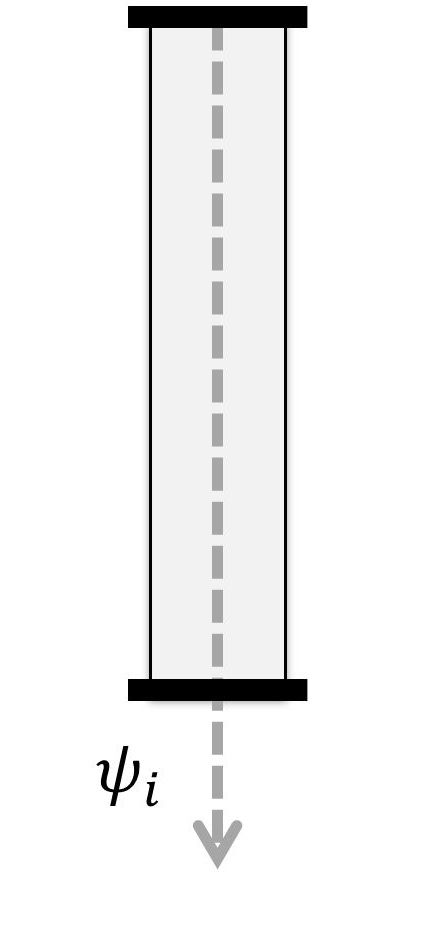}\includegraphics[scale=0.2]{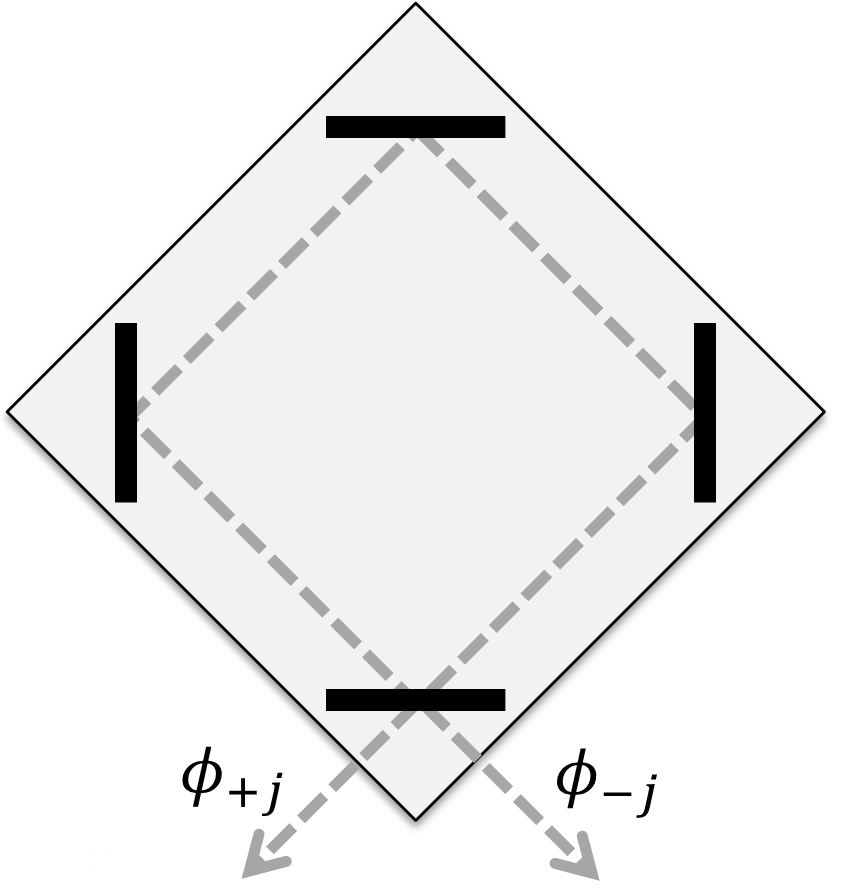}
\par\end{centering}

\caption{\label{fig:res}Linear and ring resonators}

{\footnotesize Different cavity configurations. In the linear cavity
the eigenmodes are standing waves and there is only one mode for each
frequency. In the ring cavity two counterpropagating travelling wave
modes of the same frequency can coexist and be extracted independently.}
\end{figure}

This letter introduces the idea of producing DCE in an optical ring
resonator (like a Sagnac interferometer) filled with a dielectric
subjected to a periodic modulation of the refractive index. This configuration
possesses the following advantages: (i) the pairs of photons generated
by DCE do not reflect into each other but propagate in opposite directions
and can be extracted independently (see Fig. \ref{fig:res}), remaining
therefore distinguishable; (ii) the two photon beams propagate along
the same optical path and experience the same decoherence and losses
(they share the same bath) which allows for the existence of a decoherence
free subspace (DFS) \cite{0305-4470-37-15-L04}; and (iii) the degree
of entanglement between the two beams is a mark of ZPF even at finite
temperature. 

The models for DCE in linear cavities predict that the number of photons
only grows exponentially if the rate of their creation is faster than
they decay (weak losses regime) \cite{PhysRevA.58.4147}, but even
then their growth rate is reduced by losses. For ring cavities, the
DCE can support an exponential photon generation and entanglement,
even in the strong losses regime. 

\textbf{\emph{Classical vs. quantum model. }}The classical model for
an optical cavity filled with a time dependent dielectric is basically
the same for both linear and ring resonators and can be derived directly
from the macroscopic Maxwell equations by imposing a time modulation
of the refractive index $n\equiv n(t)=n_{0}+\delta n\ h(t)$, with
$h(t)\in[-1,1]$. The resulting dynamical eq. for the amplitude of
electric field $\mathbf{E}$ of an harmonic mode is 
\begin{equation}
\ddot{\mathbf{E}}+4\dot{n}n\dot{\mathbf{E}}+(\omega^{2}+2\ddot{n}n+2\dot{n}^{2})\mathbf{E}=0\label{eq: dinamical E}
\end{equation}
and can be converted into the Hill equation \cite{magnus2004hill}
\begin{equation}
\ddot{y}=-f^{2}y,\label{eq: Hill}
\end{equation}
with $f\equiv\omega^{2}-4\dot{n}^{2}n^{2}\approx\omega_{0}^{2}(1-2\delta n\ h(t)/n_{0})$,
by introducing the variable $y=E\ e^{-n^{2}}$ and considering small
modulations of the refractive index ($\delta n/n_{0}\ll1$). 

For periodic modulations $f$ with period $T$, the formal solution
of eq. (\ref{eq: Hill}) is 
\begin{equation}
\left[\begin{array}{c}
y(t)\\
\dot{y}(t)
\end{array}\right]=S(t)\left[\begin{array}{c}
y(t=0)\\
\dot{y}(t=0)
\end{array}\right].
\end{equation}
According to Floquet's theory \cite{magnus2004hill}, to characterise
the evolution of the system at times $t=mT$ (with $m$ an integer)
it suffices to consider the Lyapunov exponent $\mu$ (i.e. the rate
of growth of the amplitude of the solution) which is computed as 
\begin{equation}
\mu=T^{-1}\cosh^{-1}\{Tr[2S(T)]\}.
\end{equation}

The solutions are bounded if $\left|Tr[2S(T)]\right|<1$, and unbounded
for $\left|Tr[2S(T)]\right|>1$, as a result of a parametric resonance
(PR). Including optical losses, the photon yield at a PR is 
\begin{equation}
N(mT)=\frac{2\epsilon_{0}n_{0}^{2}}{\hbar\omega_{0}}\left\langle E_{0}^{2}\right\rangle e^{2(\mu-\gamma)mT}
\end{equation}
where $E_{0}$ is the initial field amplitude and $\gamma$ is the
coupling of the cavity with the environment. 

At $\mathcal{T}=0$, the amount of light emitted from the ground state
of the field depends on the nature of the field considered: in the
classical model $\left\langle E_{0}^{2}\right\rangle =0$ and there
is no photon emission whereas in the quantum model $\left\langle E_{0}^{2}\right\rangle >0$
due to the ZPF, yielding an exponential amplification of the field.
But for $\mathcal{T}>0$, the difference between the classical and
quantum prediction is very small because the contribution from thermal
fluctuations present in both models is relevant and masks the contribution
from ZPF. At $\mathcal{T}>0$, the quantum character of the DCE is
hidden in the quantum correlations, calculated using a fully quantum
dynamics. This is done replacing the $\mathbf{E}$ field in eq. (\ref{eq: dinamical E})
by an operator and computing the effective Hamiltonian for each mode.
The difference between linear and ring resonators results from the
mode structure of the $\mathbf{E}$ operator determined by the geometry
of the cavity. The effects of losses and decoherence are modelled
via the coupling to a Markovian bath according to the master eq. (similar
to refs \cite{PhysRevA.58.4147,0305-4470-37-15-L04})

\begin{eqnarray}
\dot{\rho} & = & -i\left[H,\rho\right]+\frac{\gamma}{2}\underset{i,j}{\sum}\left\{ \left(\bar{n}+1\right)a_{i}\rho a_{j}^{\dagger}-\right.\nonumber \\
 &  & \left.-\left(\bar{n}+\nicefrac{1}{2}\right)\left[a_{i}^{\dagger}a_{j}\rho+\rho a_{i}^{\dagger}a_{j}\right]+\bar{n}\left(a_{i}^{\dagger}\rho a_{j}-\rho\right)\right\} .\label{eq:master}
\end{eqnarray}

\textbf{\emph{DCE in linear resonators.}}\textbf{ }For linear cavities
the $\mathbf{E}$ operator for transverse modes is 
\begin{equation}
\mathbf{E}=i\sum_{j}\xi_{j}(a_{j}-a_{j}^{\dagger})\psi_{j}(x)\mathbf{e}_{j},\label{eq:linear cavity field}
\end{equation}
where $x$ is the coordinate along the optical axis of the resonator,
$\xi_{j}(t)=\sqrt{\hbar k_{0}c|j|/2n^{2}(t)}$, $a_{j}^{\dagger}$
and $a_{j}$ are the creation and annihilation field operators (with
$[a_{i},a_{j}^{\dagger}]=\delta_{ij}$), $\mathbf{e}_{j}$ is the
polarisation unit vector and $\psi_{j}(x)=sin(k_{j}x)$ with $k_{j}=\pi j/L$
(with $j$ a positive integer).

Replacing (\ref{eq:linear cavity field}) into (\ref{eq: dinamical E})
yields the evolution eqs. for $a_{j}^{\dagger}$ and $a_{j}$, which
for each mode $\psi_{j}$ correspond to the Hamiltonian
\begin{equation}
H(t)=f(t)a_{j}^{\dagger}a_{j}+ig\left(t\right)[a_{j}^{\dagger2}-a_{j}^{2}],\label{eq:Hamiltonian linear cavity}
\end{equation}
with $f(t)=1/n(t)$ and $g(t)=\nicefrac{1}{2}d\ ln\ n(t)/dt$. From
this point all eqs. are written in natural units of the vacuum frequency,
defined by $\hbar=1$ and $k_{j}c=1$.

For Gaussian states \cite{Qlect}, the Wigner function has the form
$W(X)=\left(2\pi\right)^{-2}\det\left[\sigma\right]^{-\frac{1}{2}}\exp\{-\frac{1}{2}(\boldsymbol{X}-\overline{\boldsymbol{X}})^{T}\sigma^{-1}(\boldsymbol{X}-\overline{\boldsymbol{X}})\}$,
where $\boldsymbol{X}=\left[x,p\right]^{T}$ and $\sigma$ is a covariance
matrix defined by $\sigma=\overline{\boldsymbol{X}\boldsymbol{X}^{T}}-\overline{\boldsymbol{X}}\ \overline{\boldsymbol{X}^{T}}$.
For the Hamiltonian (\ref{eq:Hamiltonian linear cavity}), the master
eq. (\ref{eq:master}) yields the evolution eqs.
\begin{eqnarray}
\dot{\overline{\boldsymbol{X}}} & = & -\left(M_{2}^{T}+\nicefrac{\gamma}{2}\right)\overline{\boldsymbol{X}},\label{eq:dynamical eq X linear}\\
\dot{\sigma} & = & -M_{2}^{T}\sigma-\sigma M_{2}-\gamma\left(\sigma-\sigma{}_{\infty}\right),\label{eq:dynamical eq sigma linear}
\end{eqnarray}
where
\begin{equation}
M_{2}=2\left[\begin{array}{cc}
-g & f\\
-f & g
\end{array}\right],
\end{equation}
 and where $\sigma{}_{\infty}=(2\bar{n}+1)\mathrm{Id_{2}/4}$ and
$\mathrm{Id_{n}}$ is the n-dimensional identity matrix. The state
$\sigma{}_{\infty}$ describes the cavity in equilibrium with the
bath in the absence of the refractive index modulation. Notice that
$\sigma_{\infty}$ is \emph{de facto} the ground state of the cavity
at $\mathcal{T}>0$ therefore, it is simultaneously the initial state
of the cavity at $t=0$ (i.e. $\sigma_{0}=\sigma_{\infty}$) and the
asymptotic state to which the cavity decays if the modulation if turned
off.

\textbf{\emph{DCE in ring resonators}}\textbf{.} For ring cavities
the $\mathbf{E}$ operator for transverse modes is
\begin{equation}
\mathbf{E}=i\sum_{j}\xi_{j}(t)\left[a_{j}(t)\phi_{j}(x)-a_{j}^{\dagger}(t)\phi_{j}^{*}(x)\right]\mathbf{e}_{j},
\end{equation}
 where $\phi_{j}(x)=exp(ik_{j}x)$, with $k_{\pm j}=\pm2\pi j/L$
and $j$ an integer ($+j$ for light propagation along the optical
axis and $-j$ in the opposite direction). In this case, there is
a coupling between modes with symmetric wave number $j$ resulting
in the Hamiltonian \cite{PhysRevA.72.063805}
\begin{equation}
H(t)=f(t)[a_{j}^{\dagger}a_{j}+a_{-j}^{\dagger}a_{-j}]+ig\left(t\right)[a_{j}^{\dagger}a_{-j}^{\dagger}-a_{j}a_{-j}],\label{eq:Hamiltonian}
\end{equation}
Making $\boldsymbol{X}'=\Gamma\boldsymbol{X}$, with $\boldsymbol{X}=\left[x_{j},p_{j},x_{-j},p_{-j}\right]^{T}$and
\begin{equation}
\Gamma=\frac{1}{\sqrt{2}}\left[\begin{array}{cc}
\mathrm{Id_{2}} & \mathrm{Id_{2}}\\
\mathrm{Id_{2}} & \mathrm{-Id_{2}}
\end{array}\right],
\end{equation}
the evolution eqs. for $\overline{\boldsymbol{X}}$ and $\sigma$
become 
\begin{eqnarray}
\dot{\overline{\boldsymbol{X'}}} & = & -\left(M_{4}^{T}+\nicefrac{\gamma}{2}\right)\overline{\boldsymbol{X'}},\label{eq:dynamical eq X ring}\\
\dot{\sigma'} & = & -M_{4}^{T}\sigma'-\sigma'M_{4}-\nicefrac{\gamma}{2}\left(\alpha\sigma'+\sigma'\alpha-2\alpha\sigma'_{\infty}\right),\label{eq:dynamical eq sigma ring}
\end{eqnarray}
where $M'_{4}$, $\alpha$ and $\sigma'_{\infty}$ are $4\times4$
matrices with $\sigma'_{\infty}=(2\overline{n}+1)\mathrm{\alpha'/4}$
and 
\begin{equation}
M_{4}=\frac{1}{2}\left[\begin{array}{cc}
M_{2} & 0\\
0 & -M_{2}^{T}
\end{array}\right],\ \alpha'=\left[\begin{array}{cc}
\mathrm{Id_{2}} & 0\\
\mathrm{0} & \mathrm{0}
\end{array}\right].
\end{equation}

Notice that $M_{4}$, $\alpha$ and $\sigma'_{\infty}$ are block
diagonal. Then, eqs. (\ref{eq:dynamical eq X ring}) and (\ref{eq:dynamical eq sigma ring})
imply that, in terms of quadratures $\boldsymbol{X}'$, two coupled
modes of ring resonator can be decomposed into two independent modes
of a linear resonator. Moreover, one these modes is isolated from
the bath and constitutes a DFS. Hence, when a PR occurs, it grows
exponentially with $\mu$, regardless of the losses. This feature
of the ring geometry does not exist for linear cavities, where all
modes are coupled to the bath.

\textbf{\emph{Evolution of the state of the system}}. The solutions
of eqs. (\ref{eq:dynamical eq X ring}) and (\ref{eq:dynamical eq sigma ring})
for a ring resonator are
\begin{eqnarray}
\overline{\boldsymbol{X}'}(t) & = & U_{th}\overline{\boldsymbol{X'}}_{0},\\
\sigma'(t) & = & U_{th}\sigma'_{0}U_{th}^{T}+\sigma'_{p},\label{eq: dynamical equation for sigma}
\end{eqnarray}
where $\sigma'_{0}\equiv\sigma'_{\infty}$ and $\overline{\boldsymbol{X}}_{0}$
describe the initial state of the cavity, $U_{th}=e^{-\nicefrac{\gamma}{2}\alpha t}U$,
$U$ solves the eqs. in the absence of the bath and $\sigma'_{p}$
is a particular solution of eq. (\ref{eq: dynamical equation for sigma}).
In this case, $U$ and $\sigma'_{p}$ are
\begin{equation}
\begin{array}{cc}
U=\frac{1}{2}\left[\begin{array}{cc}
\mathrm{U_{+}} & \mathrm{0}\\
0 & \mathrm{U_{-}}
\end{array}\right], & \sigma'_{p}=\frac{1}{2}\left[\begin{array}{cc}
\sigma_{+} & 0\\
0 & 0
\end{array}\right]\sigma_{\infty}\end{array},
\end{equation}
where $U_{\pm}=e^{\pm G\Sigma}R_{\pm}^{-1}SR_{\pm}$ with $G=\int_{0}^{t}g\left(\tau\right)d\tau$,
$f_{0}\equiv f(t=0)$, 
\[
R_{+}=\left[\begin{array}{cc}
0 & 1\\
-f_{0} & 0
\end{array}\right],R_{-}=\left[\begin{array}{cc}
1 & 0\\
0 & f_{0}
\end{array}\right],\Sigma=\left[\begin{array}{cc}
1 & 0\\
0 & -1
\end{array}\right],
\]
and $\sigma_{+}=\gamma U_{+}\left[\int_{0}^{t}\left(U_{+}^{T}U_{+}\right)^{-1}d\tau\right]U_{+}^{T}$.
The matrix $S$ is again obtained as the solution of eq. (\ref{eq: Hill}).
The solutions of eqs. (\ref{eq:dynamical eq X linear}) and (\ref{eq:dynamical eq sigma linear})
for a linear resonator are formally identical to those of eqs. (\ref{eq:dynamical eq X ring})
and (\ref{eq:dynamical eq sigma ring}) for the degree of freedom
coupled to the bath expressed in terms of $\boldsymbol{X}'$.

\textbf{\emph{Examples of refractive index modulations.}} The impact
of the DFS is clearer by considering the two particular optical modulations:
(i) a periodic and instantaneous change of $n$ between two values,
$n_{1}$ and $n_{2}$, which is directly related to TR; and (ii) a
sinusoidal modulation, similar to more conventional setups of DCE
\cite{1402-4896-82-3-038105}. 

For a sequence of instantaneous perturbation of the refractive index
with period $T$ given by

\begin{equation}
f(t)=\begin{cases}
f_{1}, & mT<t<mT+t_{1},\\
f_{2}, & mT+t_{1}<t<mT+t_{1}+t_{2}=\left(m+1\right)T,
\end{cases}
\end{equation}

Using eqs. (\ref{eq:dynamical eq X ring}) and (\ref{eq:dynamical eq sigma ring}),
it results that 
\begin{equation}
Tr[2S(T)]=\frac{(1+f_{r})^{2}}{4f_{r}}\cos\theta_{+}-\frac{(1-f_{r})^{2}}{4f_{r}}\cos\theta_{-},\label{eq:mu}
\end{equation}
with $\theta_{\pm}=\theta_{1}\pm\theta_{2}$, $\theta_{1}=f_{1}t_{1}$,
$\theta_{2}=f_{2}t_{2}$ and $f_{r}=f_{2}/f_{1}$.

\begin{figure}
\begin{centering}
\includegraphics[scale=0.4]{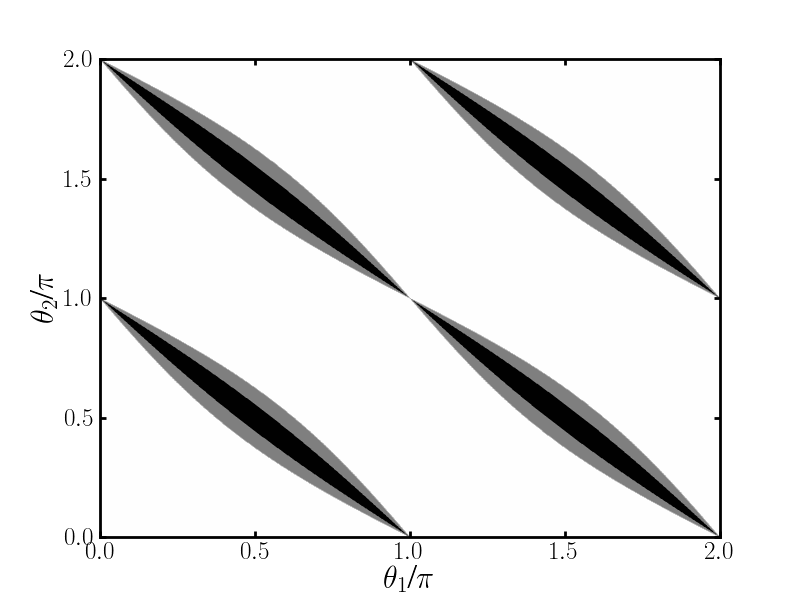}
\par\end{centering}

\caption{Phase resonance}

{\footnotesize The shaded regions are those where there is net amplification
of the field (black: $f_{r}=1.2$, grey: $f_{r}=1.5$). Maximum amplification
is obtained for $\theta_{i}=\nicefrac{\pi}{2}+k\pi$ but slight deviations
from these values still produce amplification. The allowed deviation
increases with the relative variation of the refractive index.}
\end{figure}

For $\gamma=0$, eq. (\eqref{eq:mu}) shows that the highest photon
emission rate occurs for $\theta_{i}=\nicefrac{\pi}{2}+k\pi$ (with
$k$ integer) and $\mu=T^{-1}\left|\ln f_{r}\right|$. For the same
$\theta_{i}$ but with $\gamma\neq0$, the asymptotic values of the
number of photons emitted $\mathcal{N}+1\equiv\sum_{i=1,2}(a_{i}^{\dagger}a_{i}+h.c.)/2$
and of the entanglement of the state (measured using logarithmic negativity)
at $t=mT\rightarrow\infty$ are respectively
\begin{eqnarray}
\left\langle \mathcal{N}+1\right\rangle  & \underset{mT\rightarrow\infty}{\rightarrow} & \frac{2\overline{n}+1}{4}[e^{2\mu mT}+e^{-2\eta_{-}mT}+\nonumber \\
 &  & +F_{-}^{i}(m)+F_{+}^{i}(m)]\label{eq:N_discrete}\\
E_{N}(mT) & \underset{mT\rightarrow\infty}{\rightarrow} & \max\{0,\frac{\mu mT}{\ln2}-\nonumber \\
 &  & -\log_{2}\left[\left(2\overline{n}+1\right)F_{+}^{i}(m)^{\nicefrac{1}{2}}\right]\}
\end{eqnarray}
 with $F_{\pm}^{i}(m)\equiv[1-e^{-2\eta_{\pm}mT}][1-e^{2\gamma t_{1}}+e^{2\gamma t_{1}}(1-e^{2\gamma t_{2}})f_{r}^{\pm1}]/[1-e^{2\eta_{\pm}T}]$
and $\eta_{\pm}=(\gamma\pm\mu)$.

In the case of a small sinusoidal perturbation of the refractive index,
given by $n=n_{0}+\delta n\sin\Omega t$, and for $\delta n/n_{0}\ll1$,
yields that $f^{2}\simeq[1-2(\delta n/n_{0})\sin\Omega t]/n_{0}^{2}$
and eq. (\ref{eq: Hill}) reduces to Mathieu's eq.
\begin{equation}
\ddot{y}+\left(\delta+\epsilon\sin2t\right)y=0\label{eq:Mathieu}
\end{equation}
with $t'=\Omega t/2$, $\delta=4/n_{0}^{2}\Omega^{2}$ and $\epsilon=-8\delta n/n_{0}^{3}\Omega^{2}$.
Eq. (\ref{eq:Mathieu}) exhibits PR for $\delta=m^{2}$ (with $m$
integer) or alternatively $\Omega=2/mn_{0}$, yielding $\mu=\delta n/2n_{0}^{2}.$ 

The asymptotic values of $\mathcal{N}$ and $E_{N}$ at $t=mT\rightarrow\infty$
are
\begin{eqnarray}
\left\langle \mathcal{N}+1\right\rangle  & \underset{mT\rightarrow\infty}{\rightarrow} & \frac{2\overline{n}+1}{4}[e^{2\mu mT}+e^{-2\eta_{-}mT}+\nonumber \\
 &  & +F_{-}^{s}(m)+F_{+}^{s}(m)]\label{eq:N_sinosoidal}\\
E_{N} & \underset{mT\rightarrow\infty}{\rightarrow} & \max\{0,\frac{\mu mT}{\ln2}-\nonumber \\
 &  & -\log_{2}\left[\left(2\overline{n}+1\right)F_{+}^{s}(m)^{\nicefrac{1}{2}}\right]\}
\end{eqnarray}
with $F_{\pm}^{s}(m)=[2\gamma(1-e^{2\eta_{\mp}T})(1-e^{-2\eta_{\pm}mT})]/[\eta_{\mp}(1-e^{2\eta_{\pm}T})]$.

In both cases there are two distinct contributions for $\mathcal{N}$.
The terms in (\ref{eq:N_discrete}) and (\ref{eq:N_sinosoidal}) proportional
to $e^{2\mu mT}$result from the degree of freedom associated with
the DFS, resulting in a photon emission rate of $\tau_{DFS}=2\mu$,
independently of the bath or losses. The terms proportional to $e^{-2\eta_{-}mT}$
and $F_{-}$, result from the remaining degree of freedom which is
coupled to the bath and has two regimes: (i) a strong losses regime
for $\eta_{-}>0$, when there is no exponential amplification of the
field and (ii) a weak losses regime for $\eta_{-}<0$, when photons
are produced at the rate $\tau_{th}=2(\mu-\gamma)$. This is the emission
rate for linear resonators, in which there is no DFS.

The other relevant feature of the solution is the existence of entanglement
in both regimes. In fact, since $F_{+}\left(\infty\right)$ is constant,
the linear growth term always dominates for large $m$. This means
that a long enough modulation of the medium will always result in
entanglement after an occurrence time
\begin{equation}
t_{occ}=\frac{\log_{2}\left[\left(2\overline{n}+1\right)F_{+}(\infty)^{\nicefrac{1}{2}}\right]\ln2}{\mu T}.
\end{equation}

\begin{figure}
\begin{centering}
\includegraphics[scale=0.4]{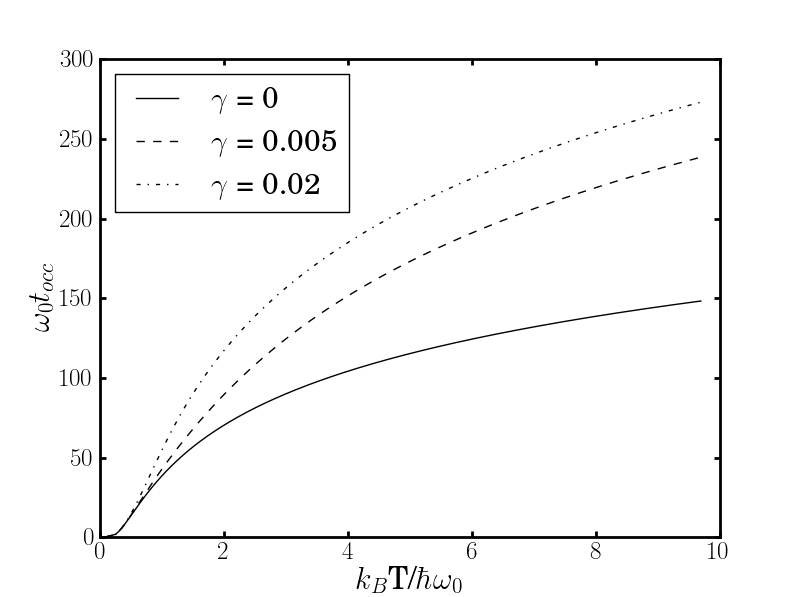}
\par\end{centering}

\caption{Occurrence time}

{\footnotesize The occurrence time as a function of the temperature
for different values of $\gamma$. The occurrence time increases with
the temperature and with $\gamma$.}
\end{figure}

It is important to note that a finite temperature does not forbid
the existence of entanglement but delays its occurrence, which enforces
the need to work with low temperatures. These results compare well
with reference \cite{0305-4470-37-15-L04}, where the author considered
two harmonic oscillators prepared in an initial two mode squeezed
state, not directly coupled but sharing a thermal bath. It was shown
that, if the initial state was sufficiently entangled, it would preserve
some amount of entanglement after any arbitrarily large time. In the
case presented here the initial state is disentangled and it is the
DCE that produces correlations until they reach a threshold imposed
by the thermal bath, after which entanglement is present. The common
factor in both cases is the existence of a DFS which shields the entanglement
from the environment.

\textbf{\emph{Conclusions. }}This letter introduces the idea of using
entanglement as a marker of the quantum nature of the DCE at finite
temperature and proposes a new experimental setting which has a DFS
and is capable of supporting and amplifying entanglement. In more
detail, it was shown that for a resonant modulation of the refractive
index of the dielectric medium in a ring resonator it is always possible
to amplify radiation, regardless of the loss regime. Furthermore,
it was verified that the existence of a decoherence free subspace
allows quantum correlations to survive, making entanglement viable
even at high temperatures. Entanglement is a distinctive character
of a quantum effect which can be used to sort out the DCE from classical
parametric amplification of thermal fluctuations. Hopefully these
results will pave way to the observation of the DCE and entanglement
at finite temperatures.

\thanks{Acknowledgements }

A. G. acknowledges the support of the Casimir network of the European
Science Foundation.

\bibliographystyle{unsrtnat}

\end{document}